\documentclass[runningheads]{llncs}
\usepackage{amsmath}
\usepackage{amsfonts}
\usepackage[english]{babel}
\usepackage{amssymb}
\usepackage[sort,compress]{cite}

\usepackage{graphicx}

\usepackage{tikz}
\usepackage{pgfplots}
\pgfplotsset{ytick scale label code/.code={$1 \times 10^{#1}$}, 
	xtick scale label code/.code={$1 \times 10^{#1}$},
}
\pgfkeys{/pgf/number format/.cd, std,
	precision=2,
	1000 sep={{{{,}}}},
	min exponent for 1000 sep=4,
	sci generic={mantisse sep=\times,exponent={10^{#1}}}}
\begin{document}
\title{Scale-free network generation model with addition and deletion of nodes based on triadic closure mechanism\thanks{The work was supported by the Russian Science Foundation, project 23-21-00148.}}
\titlerunning{Scale-free network generation model}
%
\author{Sergei Sidorov\inst{1}\orcidID{0000-0003-4047-8239} \and
Sergei Mironov\inst{1}\orcidID{0000-0003-3699-5006} \and
Timofei D. Emelianov\inst{1}}
\authorrunning{S. Sidorov et al.}
%
\institute{Saratov State University, Saratov 410012, RUSSIA
\email{sidorovsp@sgu.ru}}
\maketitle              
\begin{abstract}
Many real systems exhibit the processes of growth and shrink. In this paper, we propose a network evolution model based on the simultaneous application of both node addition and deletion rules. To obtain a higher clustering that is present in real social networks, the model employs the triadic formation step at each iteration. The results show that the degree distribution in the networks generated based on this model follows a power law.

\end{abstract}
\section{Introduction}

Studying the principles and mechanisms of network generation is a crucial area of research in the field of complex networks theory. Researchers primarily grapple with the challenge of devising algorithms that can effectively simulate network evolution, resulting in synthetic graphs that possess certain characteristics akin to real-world complex networks. As a property that it is desirable to reproduce in the process of generating a network, the power law for the degree distribution is often considered.Indeed, many real networks have the scale-free property, that is, the degree distribution in such networks follows a power law with a quite small exponent.

Most real networks experience growth by incorporating new nodes. This growth is evident in social networks as they attract new users, in the expansion of the World Wide Web through the addition of new pages and sites, and in citation networks that expand with the emergence of new publications citing previously published ones. Therefore, the first network generation models were focused on modeling the process of network growth, employing algorithms based on preferential attachment mechanisms. In the last two decades, multiple models have been developed that further refine these concepts. However, almost all the works were devoted to the modeling of growing networks, while the study of contraction and shrinking mechanisms turned out to be practically beyond the area of interest of scientists. This can be explained by the following. Firstly, most real networks tend to grow, and the processes of reduction have only recently gained attention in various real-world systems. Secondly, modeling shrinking processes has proven to be challenging, as they are less intuitive and do not lend themselves to being described by applying growth mechanisms in reverse chronological order.

Meanwhile, it is important to note that many systems and networks also undergo contraction as they grow, which involves the removal of elements or links in addition to the incorporation of new nodes or subsystems. In fact, these processes of node addition and removal often take place concurrently. However, the existing scientific literature lacks sufficient models that effectively capture the evolution of such networks. One model that addresses this gap is presented in \cite{pub.1032738127}. This model incorporates node deletion and duplication mechanisms, and the study demonstrates that the resulting graphs exhibit significant properties observed in protein interaction networks. Furthermore, the paper  \cite{PhysRevE.74.036121} introduces a model in which new nodes are added to the network while random nodes are deleted with a fixed probability. The study shows that the rate of node addition must exceed the rate of deletion in order to maintain the scale-free nature of the evolving network. Additionally, the work \cite{PhysRevE.65.057102} utilizes the list of active nodes, deleting from which deactivates the node and excludes the node from the process of network growth in subsequent iterations.

The paper \cite{Ben-Naim_2007} proposes a model for growing network generation with both addition and removal of nodes. The model uses two independent steps at each iteration. Firstly, with rate $0<r<1$, a node is attached to the graph and this node joins to a randomly selected node of the network. Secondly, a randomly selected vertex is removed, and one of its neighbors obtains the links of its immediate descendants. The paper \cite{DENG2007714} examines a network growth mechanism which includes addition and removal of nodes. Specifically, the authors focus on analyzing the impact of node removal on the overall structure of the graph. It is shown that the use of the node removal rule leads to a transition from a scale-free distribution to an exponential type distribution.
Another mechanism for generating networks based on edge removal was proposed in \cite{Deijfen288753}, in which the authors investigated a model that involves the simultaneous use of the preferential attachment step and the step of removing a random edge from the network.

A recent paper \cite{10.1371/journal.pone.0223480} introduced a novel approach for constructing efficient networks by merging vertices. This model demonstrated the ability to accurately simulate networks with a power-law degree distribution. The concept of merging blocks within networks has also been explored in previous research \cite{PhysRevE.72.046116}. Additionally, a recent paper \cite{ZENG2023151} proposed a model for simulating the evolution of networks through the state transition of vertices between online and hidden states, using a birth and death process.

In this article, we propose a new model for generating complex networks whose evolution involves the simultaneous use of growth and contraction mechanisms. Growth in the model will be provided by adding one new vertex at each iteration. The contraction will occur by merging two randomly chosen vertices. The result of the merging will be a vertex whose neighbors are the neighbors of these two vertices.

The paper is organized as follows. First, in Section \ref{sec:290yfhhhjhs}, we will consider a model in which at each iteration one node is added to the network, one node removed from the graph, while the triadic closure mechanism will not be used. We will show that the limiting degree distribution is not depend on the distribution of the initial network. Then, in Section \ref{sec:0qvjojvojnbo}, we propose a model that complements the model in Section \ref{sec:290yfhhhjhs} with the ability to form triads by drawing an edge between the neighbors of the merging nodes.

The network growth step in our model is implemented using a triadic closure mechanism \cite{PhysRevE.65.026107}. The use of the triadic closure as a network growth mechanism in this study is motivated by the fact that the simulated networks exhibit a high clustering coefficient typically observed in real networks \cite{PhysRevE.68.036122}, \cite{PhysRevE.84.066117}.

One of the problems that arises when analyzing complex networks is to coarse the graph through various methods, including dimension reduction, removing vertices or edges, so that the essential properties or structural features of the original system are preserved \cite{s40324-021-00282-x,10.1145/3447548.3467256}. We would like to note that this study addresses a different problem related to the search for realistic mechanisms for the evolution of complex networks, and not to the development of graph coarsening techniques.


\section{One-in and one-out model}\label{sec:290yfhhhjhs}

\subsection{The model}

Let us describe the model formally. Let at the initial time the network consists of $n$ vertices and $m$ edges. We assume that the vertices are numbered by integers $V_n:=\{1,2,\ldots,n\}$.

The evolution of the networks involves two steps at each iteration:
\begin{itemize}
    \item 
    (\textit{Growth}). At each iteration $t=n+1,\ldots$ a new node $t$ is added to the network, i.e. at iteration $t$, the set of vertices $V_n:=\{1,2,\ldots,t\}$. This new vertex is connected by an edge to one of the already existing vertices in the network. The vertex to which a new vertex is attached is chosen randomly among all nodes with non-zero degree (the probability of choosing a vertex is $\frac 1n$).
    \item
    (\textit{Contraction}). In addition, at each iteration two vertices are randomly selected: a new vertex is formed, the identifier of which will be equal to the number of the first vertex, the second vertex is removed (i.e. takes degree 0), and all its neighbors are attached to the first vertex. If some neighbor of the second vertex was also a neighbor of the first, then the edge is not duplicated. Thus, the degree of the first vertex is increased by the degree of the second vertex, excluding the number of neighbors of the second node that are also neighbors of the first one.
\end{itemize}

Since the new vertex is joined with one edge, the total number of edges is increasing by 1 from iteration to iteration. Moreover, since at each iteration one new vertex is added to the network, but one vertex is removed as a result of the merging, the total number of nodes does not change from iteration to iteration. The edge between the merging vertices disappears if it exists. 

Simultaneous use of the mechanism of growth and reduction of the network leads to the fact that the size of the network remain unchanged in the process of evolution, but its edge density is increasing. However, as a result of repeated application of the steps of the algorithm, the degree of each of the vertices may change, as well as the distribution of the degrees of the network. The subject of this paper is the analysis of the degree distribution evolution in networks generated by this model.



\subsection{Degree distribution analysis}

In this section, we will show that the limiting degree distribution in the networks generated by this model will converge to a power law, whatever the network's degree distribution is at the initial time.

Denote by $q_k(t)$ the fraction of network nodes that have degree $k$ at time $t$. Let $q_k=\lim_{t\to \infty}q_k(t)$ be the limit value of the probability that a randomly selected network node has degree $k$.

Note that the degree of a node can change
\begin{itemize}
	\item
	 if a new node chooses it (with probability $p_1=\frac 1n$), 
	 \item
	 if this node is chosen as the second node when two randomly selected network nodes are merged (probability $p_2=\frac 1n$), 
	 \item
	 if the node is chosen as the first node in the merge (probability $p_3=\frac 1n$).
\end{itemize}

Let us first study how the proportion of nodes with degree 1 changes at iteration $t+1$. First, a node can have degree 1 at iteration $t$ and none of the possible situations leading to a change in its degree has occurred. The probability of such an outcome is $1-p_1-p_2-p_3$. Secondly, at each iteration a new node is added with degree 1 (with one node removed at the step of merging nodes, so the number of nodes in the network remains unchanged and equal to $n$). Therefore, the fraction of nodes with degree 1 increases by $\frac 1n$ at each iteration. Thus, we get  the equation
\[
q_i(t+1)=q_1(t)(1-p_1-p_2-p_3)+\frac 1n,
\]
whence for $t\to \infty$ we get $q_1=\frac 13$.

Further, the proportion of nodes that have degree 2 can change at iteration $t+1$ in two cases: first, if the node had degree 1 at iteration $t$ and a new node has joined it; second, if a node had degree 1, it was chosen as the first node in the merging step, and the degree of the second node turned out to be 1. In addition, a node can have degree 2 at iteration $t$ and none of the possible situations leading to a change in its degree did not occur (its probability is equal to $1-p_1-p_2-p_3$). Thus we have
\[
q_2(t+1)=q_2(t)(1-p_1-p_2-p_3)+q_1(t)_1p_1+p_2q_1^2(t),
\]
whence for $t\to \infty$ we get $q_2=\frac 4{27}$, taking into account $q_1=\frac 13$.

In general, the share of vertices that have degree $k$ may change at iteration $t+1$ in two situations. First, if the vertex with degree $k-1$ at iteration $t$ joins a new node. Second, if a vertex with degree $i$ is chosen as the first node in the merging step, while the degree of the second vertex is equal to $k-i$. Also, the share of nodes with degree $k$ at iteration $t$ may do not change if none of the possible situations has occurred, i.e. with the probability $1-p_1-p_2-p_3$). Thus we have
\[
q_k(t+1)=q_k(t)(1-p_1-p_2-p_3)+q_{k-1}(t)_1p_1+p_2\sum_{i=1}^{k-1}q_i(t)q_{k-i}(t),
\]  
and we obtain the limit value as follows
\begin{equation}\label{eq:91fjo1fjd}
    q_k=\frac 13\left(q_{k-1}+\sum_{i=1}^{k-1}q_iq_{k-i}\right).
\end{equation}
It can be shown that the sequence $(q_k)_{k\geq 1}$ obtained from \eqref{eq:91fjo1fjd} follows the power law with exponent $-\frac 32$ (see Fig. \ref{fig:m3_ravnover} (a)).
    

\subsection{Empirical results}

We carry out experiments for each type of initial network. For the initial graph consisting of $n=50$k vertices 1M steps are repeated. Let $N_1$ and $N_2$ be the two nodes which are chosen randomly with equal probability. Each neighbor of the first node $N_1$ is connected to the second node $N_2$ unless they are already neighbors. Next, all connections between $N_1$ and its neighbors are deleted, and for each of them $N_1$ is connected to another random vertex in a graph selected randomly with equal probability. To average the results, 100 independent simulations were performed and the resulting distributions are averaged upon them. These distributions are obtained for both initial and modified graphs after the final rearrangement of edges.

\begin{figure}[!ht]
    \centering\small
    \begin{minipage}{0.475\linewidth}
        \centering
        \begin{tikzpicture}\footnotesize
            \begin{axis}[height = 0.8\linewidth, width=\linewidth,
                xmin=0,
                xmax=3.1,
                tick align = {outside},
                xlabel={$\log k$},
                legend style = {cells = {anchor=west}, nodes = {scale=0.75}}, legend pos=north east
                ]
                \addplot[blue, only marks, mark=*, mark options={scale=0.35}] table[x=logi,y=logdi]{Formulae.txt};
                \addlegendentry{$\log q_k$}
                \addplot[domain=0:5, red, smooth, thick] {-1.5 * x - 0.4};
                \addlegendentry{$-1.5\log k+C$}
                
            \end{axis}
        \end{tikzpicture}
        
        (\textit{a})
    \end{minipage}
    \begin{minipage}{0.475\linewidth}
	\centering
	\begin{tikzpicture}\footnotesize
		\begin{axis}[height = 0.8\linewidth, width=\linewidth,
			xmin=0,
			xmax=3.1,
			tick align = {outside},
			                ymax=5.9,
			xlabel={$\log(k)$},
			legend style = {cells = {anchor=west}, nodes = {scale=0.75}}, legend pos=north east
			]
			\addplot[blue, only marks, mark=*, mark options={scale=0.35}] table[x=logi,y=logdi]{BR-Ravnom1.txt};
			\addlegendentry{initial distr.}
			\addplot[ violet, only marks, mark=*, mark options={scale=0.35}] table[x=logi,y=logdi]{BR-Ravnom2.txt};
			\addlegendentry{final distr.}
			\addplot[domain=0:5, red, smooth, thick] {-1.5 * x + 4.3};
			\addlegendentry{$-1.5\log k+C$}
			
		\end{axis}
	\end{tikzpicture}
	
	(\textit{b})	
\end{minipage}

\vspace{1em}
    \begin{minipage}
	{0.475\linewidth}
	\centering
	\begin{tikzpicture}\footnotesize
		\begin{axis}[height = 0.8\linewidth, width=\linewidth,
			xmin=0,
			xmax=3.1,
			tick align = {outside},
			                ymax=5.9,
			xlabel={$\log(k)$},
			legend style = {cells = {anchor=west}, nodes = {scale=0.75}}, legend pos=north east
			]
			\addplot[blue, only marks, mark=*, mark options={scale=0.35}] table[x=logi,y=logdi]{BR-ER1.txt};
			\addlegendentry{initial distr.}
			\addplot[ violet, only marks, mark=*, mark options={scale=0.35}] table[x=logi,y=logdi]{BR-ER2.txt};
			\addlegendentry{final distr.}
			\addplot[domain=0:5, red, smooth, thick] {-1.5 * x + 4.3};
			\addlegendentry{$-1.5\log k+C$}
			
		\end{axis}
	\end{tikzpicture}
	
	(\textit{c})	
\end{minipage}
    \begin{minipage}{0.475\linewidth}
	\centering
	\begin{tikzpicture}\footnotesize
		\begin{axis}[height = 0.8\linewidth, width=\linewidth,
			xmin=0,
			xmax=3.1,
			tick align = {outside},
			              ymax=5.9,
			xlabel={$\log(k)$},
			legend style = {cells = {anchor=west}, nodes = {scale=0.75}}, legend pos=north east
			]
			\addplot[blue, only marks, mark=*, mark options={scale=0.35}] table[x=logi,y=logdi]{BR-BA1.txt};
			\addlegendentry{initial}
			\addplot[ violet, only marks, mark=*, mark options={scale=0.35}] table[x=logi,y=logdi]{BR-BA2.txt};
			\addlegendentry{final}
			\addplot[domain=0:5, red, smooth, thick] {-1.5 * x + 4.3};
			\addlegendentry{$-1.5\log k+C$}
			
		\end{axis}
	\end{tikzpicture}
	
	(\textit{d})	
\end{minipage}

    \caption{(\textit{a}) Probability distribution according to Eq. \eqref{eq:91fjo1fjd}; (\textit{b}) The degree distribution in the graph in the first series of experiments, averaged over 100 simulations. The initial graph consisted of $n=50,000$ vertices and was built as a growing network; (\textit{c}) The degree distribution obtained in the second series of experiments. The initial network is the Erd\H{o}s-R\'{e}nyi random graph with $n=50000$ vertices constructed with the parameter $p=0.0001$; (\textit{d}) The average degree distribution in the graphs obtained in the third series of experiments. The initial networks are the Barab\'{a}si--Albert graphs with $n=50000$ vertices, constructed with the parameter $m=3$}\label{fig:m3_ravnover}
\end{figure}
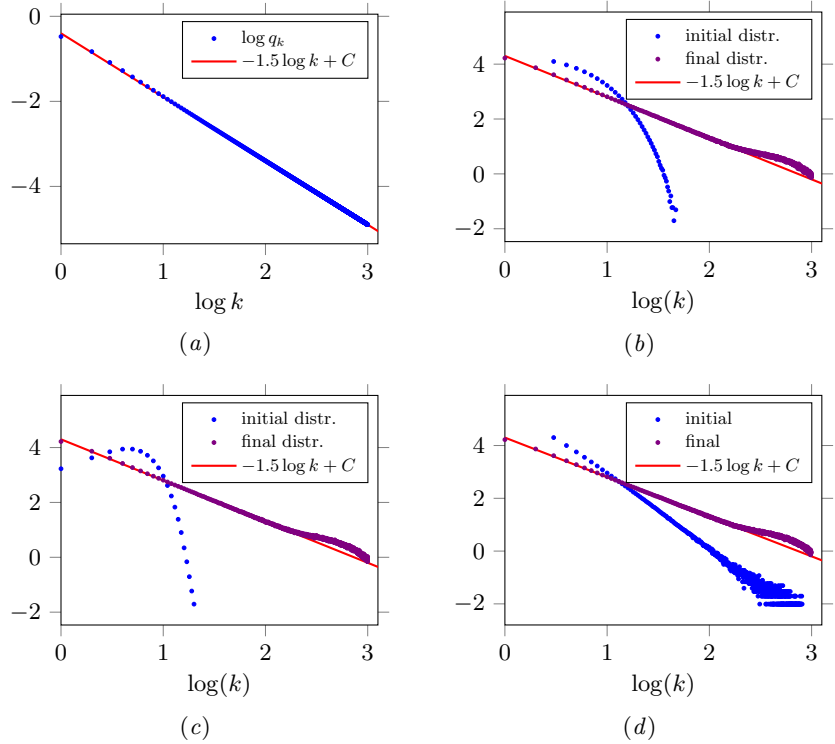

\textbf{Experiment 1.} First we define a graph generation process. Let $m$ be the size of a complete graph. Every growth step $m$ nodes are selected with equal probability. Then a new node is added, which is immediately connected to the selected nodes in the graph. In Figure~\ref{fig:m3_ravnover}\, (b) we can see the distribution of degrees in the initial graph on log-log scale (blue) and the degree distribution after application of the model after 1M repeated steps (purple). We can see that degrees of nodes after graph modifications follow the power-law with an exponent $\gamma \approx 1.5$.

\textbf{Experiment 2.} Here before the application of the model the graph is obtained according to the Erd\H{o}s-R\'{e}nyi model with the parameter $p=0.0001$. In Figure~\ref{fig:m3_ravnover} (c) we can see the averaged degree distributions. Similarly, the initial distribution is blue, and purple shows degree distribution after the repeated application of the model. The results resemble what was obtained in Experiment 1, and the distribution also follows the power law.

\textbf{Experiment 3.} The initial graph is created using the Barab\'{a}si-Albert model with the parameter $m=3$. Figure~\ref{fig:m3_ravnover}\ (d) shows the results of the simulation. Despite the fact that the initial distribution was fairly different from ones in other experiments, the final result is similar to the final results in other experiments.

\section{The one-in and one-out model with triadic closure step}\label{sec:0qvjojvojnbo}

\subsection{The model description}

The evolution of the networks involves three steps at each iteration:
\begin{itemize}
    \item 
    (\textit{Growth}). During each iteration starting from $t=n+1$, a new node $t$ is introduced to the network, which means that at iteration $t$, the set of vertices becomes $V_n:=\{1,2,\ldots,t\}$. This freshly added node is connected to one of the existing vertices in the network. The selection process for the vertex to which the new node is attached is completely random, involving all nodes that have a degree greater than zero. The probability of selecting a vertex is $\frac 1n$.
    \item
    (\textit{Contraction}). Furthermore, during each iteration, two vertices are chosen at random. The first vertex will serve as the identifier for the newly formed node, while the second vertex is removed from the network and its degree becomes zero. Additionally, all the neighbors of the second vertex are attached to the first vertex. If any of the second vertex's neighbors were already neighbors of the first vertex, the edge is not duplicated. As a result, the degree of the first vertex increases by the degree of the second vertex, excluding the number of neighbors that both the first and second nodes share in common.
    \item
    (\textit{Triad formation}). 
    Among the neighbors of each of the two nodes undergoing merging in the previous step, two nodes are selected uniformly at random (one among the neighbors of the first, the second among the neighbors of the second node). An edge is drawn between these two vertices with a given probability $p$(the model parameter).
\end{itemize}

\subsection{The degree distribution analysis}

In each experiment, 500k iterations were applied to the initial graph of 50k vertices, at each of which
\begin{enumerate}
     \item
From the entire set of vertices, two random vertices $u$ and $v$ were chosen with equal probability.
     \item
with probability $p$, an edge was drawn between some vertices $u'$ and $v'$, where $u'$ and $v'$ are two non-adjacent vertices that are neighbors of $u$ and $v$, respectively. In this case, the vertices $u'$ and $v'$ are chosen with equal probability.
     \item
neighboring vertices $u$ that are not neighbors of $v$ were connected by an edge to $v$.
     \item
all edges incident with $u$ were removed from the graph
     \item
an edge was drawn from $u$ to a random vertex of the graph. The choice of a vertex to join was carried out with equal probability.
\end{enumerate}

For the graph obtained as a result of applying the model, the distribution of vertex degrees was obtained.

For the initial graph constructed using the Barabási-Albert model, the proposed model was used. Four series of experiments were carried out for different values of the parameter $p$ of the model.

In each series, 100 experiments were carried out and the degree distribution was averaged.

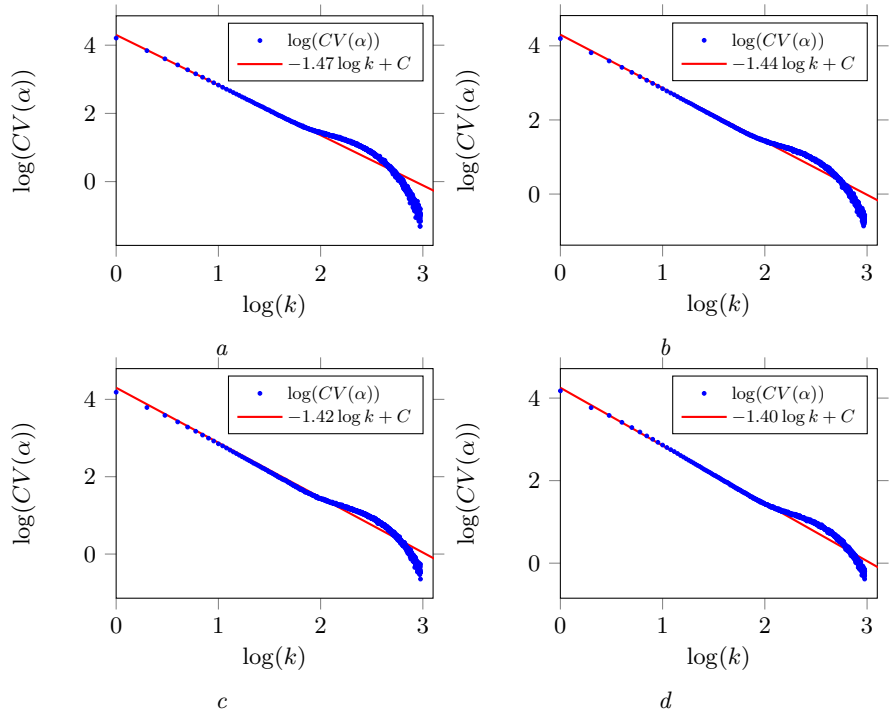
\begin{figure}[!ht]

    \begin{minipage}{0.475\linewidth}
    \centering
    \begin{tikzpicture}\footnotesize
        \begin{axis}[height = 0.8\linewidth, width=\linewidth,
            xmin=0,
            xmax=3.1,
            tick align = {outside},
            xlabel={$\log(k)$},
            ylabel={$\log(CV(\alpha))$},
            legend style = {cells = {anchor=west}, nodes = {scale=0.75}}, legend pos=north east
            ]
            \addplot[blue, only marks, mark=*, mark options={scale=0.35}] table[x=logi,y=logdi]{BAshrinkNR-BA10_degs_m3_g0.25_100000_102.inp};
            \addlegendentry{$\log(CV(\alpha))$}
            \addplot[domain=0:5, red, smooth, thick] {-1.47 * x + 4.3};
            \addlegendentry{$-1.47\log k+C$}
            
        \end{axis}
    \end{tikzpicture}
    
    \textit{a}	
\end{minipage}
 \begin{minipage}{0.475\linewidth}
    \centering
    \begin{tikzpicture}\footnotesize
        \begin{axis}[height = 0.8\linewidth, width=\linewidth,
            xmin=0,
            xmax=3.1,
            tick align = {outside},
            xlabel={$\log(k)$},
            ylabel={$\log(CV(\alpha))$},
            legend style = {cells = {anchor=west}, nodes = {scale=0.75}}, legend pos=north east
            ]
            \addplot[blue, only marks, mark=*, mark options={scale=0.35}] table[x=logi,y=logdi]{BAshrinkNR-BA10_degs_m3_g0.5_100000_102.inp};
            \addlegendentry{$\log(CV(\alpha))$}
            \addplot[domain=0:5, red, smooth, thick] {-1.44 * x + 4.3};
            \addlegendentry{$-1.44\log k+C$}
            
        \end{axis}
    \end{tikzpicture}
    
    \textit{b}	
\end{minipage}

\begin{minipage}{0.475\linewidth}
    \centering
    \begin{tikzpicture}\footnotesize
        \begin{axis}[height = 0.8\linewidth, width=\linewidth,
            xmin=0,
            xmax=3.1,
            tick align = {outside},
            xlabel={$\log(k)$},
            ylabel={$\log(CV(\alpha))$},
            legend style = {cells = {anchor=west}, nodes = {scale=0.75}}, legend pos=north east
            ]
            \addplot[blue, only marks, mark=*, mark options={scale=0.35}] table[x=logi,y=logdi]{BAshrinkNR-BA10_degs_m3_g0.75_100000_102.inp};
            \addlegendentry{$\log(CV(\alpha))$}
            \addplot[domain=0:5, red, smooth, thick] {-1.42 * x + 4.3};
            \addlegendentry{$-1.42\log k+C$}
            
        \end{axis}
    \end{tikzpicture}
    
    \textit{c}	
\end{minipage}
\begin{minipage}{0.475\linewidth}
    \centering
    \begin{tikzpicture}\footnotesize
        \begin{axis}[height = 0.8\linewidth, width=\linewidth,
            xmin=0,
            xmax=3.1,
            tick align = {outside},
            xlabel={$\log(k)$},
            ylabel={$\log(CV(\alpha))$},
            legend style = {cells = {anchor=west}, nodes = {scale=0.75}}, legend pos=north east
            ]
            \addplot[blue, only marks, mark=*, mark options={scale=0.35}] table[x=logi,y=logdi]{BAshrinkNR-BA10_degs_m3_g1.0_100000_102.inp};
            \addlegendentry{$\log(CV(\alpha))$}
            \addplot[domain=0:5, red, smooth, thick] {-1.4 * x + 4.25};
            \addlegendentry{$-1.40\log k+C$}
            
        \end{axis}
    \end{tikzpicture}
    
    \textit{d}	
\end{minipage}

    \caption{Degree distribution in the graphs obtained as a result of applying 500k iterations with parameter (a) $p=0.25$, (b) $p=0.5$, (c) $p=0.75$, (d) $p=1.0$ }\label{fig:m3_ravnover_2}
\end{figure}

The initial graph was a graph constructed using the Barab\'{a}si-Albert model with the parameter $m=3$. The results of the experiments are presented in Figure~\ref{fig:m3_ravnover_2}. The average distribution of graph degrees after 500k model steps is displayed in blue.

\section{Conclusion}

Many real systems exhibit the processes of growth and shrink. In this paper, we propose a network evolution model based on triadic closure, which performs both node addition and deletion. The resulting degree distributions in the networks generated with the usage of this model follow a power law.


\bibliographystyle{splncs04}
\bibliography{references}

\end{document}